\documentclass[11pt,letterpaper]{article}
\usepackage[margin=1in]{geometry}
\usepackage[T1]{fontenc}
\usepackage{amsmath,amssymb,amsthm}
\usepackage{xcolor}
\usepackage{hyperref}
\hypersetup{
  colorlinks,
  linkcolor={red!50!black},
  citecolor={blue!50!black},
  urlcolor={blue!80!black}
}
\usepackage[nameinlink]{cleveref}

\newtheorem{theorem}{Theorem}
\newtheorem{lemma}[theorem]{Lemma}
\newcommand{\E}{\mathbb{E}}
\newcommand{\OPT}{\mathrm{OPT}}
\newcommand{\ALG}{\mathrm{ALG}}

\title{The Optimal Competitive Ratio for Subadditive Online Combinatorial Auctions is Less Than 1/2}
\author{%
  Andr\'es Cristi\thanks{\texttt{\{andres.cristi, jiechen.zhang\}@epfl.ch}}
  \qquad Jiechen Zhang\footnotemark[1] \\
  EPFL
}
\date{}

\begin{document}
\maketitle

\begin{abstract}
We construct a four-buyer online combinatorial auction with independent
subadditive valuations and a fixed, known arrival order in which every
online rule earns strictly less than $1/2$ of the expected offline optimum.
A common partition, revealed at the third arrival, gives the first
three buyers one unit of welfare per item. If the rule uses three
distinct partitions, either an item allocated early contributes zero or
the fourth buyer's expected value for the remaining bundle is one less
than its size. We compute the optimal online and offline welfare exactly.
We also prove a $1/2$ guarantee for instances with at most three buyers.
\end{abstract}

\section{Introduction}

Let $M$ be a finite set of indivisible items. Buyers arrive in a fixed,
known order with valuations $v_i:2^M\to\mathbb{R}_{\geq 0}$ drawn
independently from known distributions. On buyer $i$'s arrival, the rule
sees $v_i$ and irrevocably allocates a bundle $S_i$ of remaining items.
Its welfare is $\ALG=\sum_i v_i(S_i)$. The prophet knows all valuations
and maximizes welfare over disjoint bundles, obtaining $\OPT$.
Valuations are monotone and satisfy $v(\varnothing)=0$. Subadditivity
means $v(S\cup T)\leq v(S)+v(T)$ for all $S,T\subseteq M$.

For one item, the prophet inequality of Krengel, Sucheston, and
Garling~\cite{KS77,KS78} guarantees $1/2$ of the expected maximum.
Samuel-Cahn~\cite{SC84} showed that a single threshold achieves this
guarantee. Feldman, Gravin, and Lucier~\cite{FGL} extended the $1/2$
guarantee to XOS valuations using static item prices. An XOS valuation
is a maximum of finitely many nonnegative additive valuations.

For subadditive valuations, Correa and Cristi~\cite{CC} proved a
$1/6$ guarantee. For any $\varepsilon>0$, MacRury et al.~\cite{MNPSSS}
gave a demand-query algorithm with guarantee $1/(6+\varepsilon)$,
improved to $1/(60/11+\varepsilon)$ for i.i.d.\ valuations. Its expected
running time is polynomial in the number of buyers and items, total
support size, and $1/\varepsilon$. For valuations with finite support,
Kroll et al.~\cite[Theorem~1.1]{KMRS} gave an expected-polynomial-time
demand-query algorithm with guarantee $1/4-\varepsilon$ for every
$\varepsilon>0$ against the stronger ex-ante configuration LP benchmark.
Taking $\varepsilon\downarrow 0$ gives a $1/4$ guarantee for the optimal
online rule. We show that $1/2$ can fail even against the expected
offline optimum.

\begin{theorem}\label{thm:main}
For some constant $\varepsilon>0$, there is a four-buyer instance with
independent, monotone, subadditive valuations and a fixed, known arrival
order in which every online allocation rule satisfies
\[
  \frac{\E[\ALG]}{\E[\OPT]}\leq\frac12-\varepsilon.
\]
\end{theorem}

\Cref{sec:3-buyers} gives an online rule that obtains at least $1/2$ of
the expected offline optimum with at most three buyers. Thus four is
the smallest number of buyers for which a $1/2$ guarantee can fail.

\paragraph{A stronger upper bound.}
A construction with a growing number of buyers gives an upper bound of
approximately $0.444211$, strictly below $4/9$. The proof and Lean
formalization are available in the
\href{https://github.com/GO-EPFL/subadditive-online-combinatorial-auctions}{GitHub repository}.

\section{The instance}

For integers $a,b\geq 2$, let $C=\{1,\ldots,a\}\times\{1,\ldots,b\}$ be
an $a\times b$ grid. An item is a labeling $x:C\to\{1,2,3\}$ that uses
a label other than $1$ in at most two cells. Let $M$ be the set of all
such labelings. Counting zero, one, and two cells with labels other
than $1$ gives
\[
  m=|M|=1+2ab+4\binom{ab}{2}=1+2(ab)^2.
\]
Each cell $c$ partitions $M$ into $A_i(c)=\{x\in M:x(c)=i\}$ for $i=1,2,3$.

\paragraph{The first three buyers.}
Buyer 1 may use any cell. Buyer 2 may use any cell in a uniformly
random row $I$. Buyer 3 may use any cell in an independently and
uniformly drawn column $J$. If $C_i$ is the set of cells available to
buyer $i$, her valuation is
\[
  v_i(S)=\max_{c\in C_i}|S\cap A_i(c)|.
\]
Each of these valuations is XOS. All three buyers can use $(I,J)$,
so allocating $A_i(I,J)$ to each buyer $i$ earns welfare $m$.

\paragraph{The fourth buyer.}
For an ordered triple $e=(e_1,e_2,e_3)$ of distinct cells, call
$B(e)=\{x\in M:x(e_i)\ne i\text{ for }i=1,2,3\}$ a \emph{block}.
The fourth buyer's expected valuation is
\[
  g(S)=|S|-d(S),\qquad
  d(S)=\mathbf{1}\{S\text{ contains a block}\}.
\]
Fix $0<q<1/m$. Independently of the first three buyers, buyer 4 has
valuation $g/q$ with probability $q$ and valuation zero otherwise.

\begin{lemma}\label{lem:intersection}
Every two blocks intersect, the function $g$ is monotone
and subadditive, and $g(\varnothing)=0$.
\end{lemma}
\begin{proof}
The all-$1$ labeling can violate the constraints of two blocks only at
the first cells of their triples. At each such cell at most two labels
are forbidden. Choose an allowed label there and leave all other labels
equal to $1$. This item belongs to both blocks, even if they coincide.

Thus every block is nonempty and $g(\varnothing)=0$. Adding an item
increases $d$ by at most one, so $g$ is monotone. For disjoint $S,T$ at
most one contains a block, giving $d(S)+d(T)\leq d(S\cup T)$ and hence
$g(S\cup T)\leq g(S)+g(T)$. For arbitrary $S,T$, the disjoint case and
monotonicity give $g(S\cup T)\leq g(S)+g(T\setminus S)\leq g(S)+g(T)$.
\end{proof}

\section{The optimal offline and online welfare}

\paragraph{The prophet.}
When buyer 4 has zero valuation, the common cell $(I,J)$ yields
welfare $m$, which is optimal since each item contributes at most one.
When her valuation is $g/q$, giving her $M$ yields welfare $(m-1)/q$.
This allocation is optimal because removing any one item $x$ from $M$
leaves a block. Indeed, choose a cell $e_1$ where $x(e_1)=1$ and complete
it to a triple $e$ of distinct cells. Then $x\notin B(e)$, so
$g(M\setminus\{x\})=m-2$. By monotonicity, giving buyer 4 any proper
subset lowers her value by at least $1/q>m$, while the first three
buyers can gain at most $m$. Averaging over buyer 4's valuation gives
\begin{equation}\label{eq:prophet}
  \E[\OPT]=(1-q)m+(m-1)=2m-1-qm.
\end{equation}

\paragraph{The online upper bound.}
After the rule assigns $S_i$ to buyer $i\leq 3$, choose $c_i\in C_i$ with
$v_i(S_i)=|S_i\cap A_i(c_i)|$, breaking ties by a fixed order. Let $R$
be the items remaining after buyer 3. Conditional on the history through
buyer 3, including the rule's random choices, buyer 4 has expected
value $g(R)$ for $R$. By monotonicity, conditional expected welfare is
at most
\[
  \sum_{i=1}^3 |S_i\cap A_i(c_i)|+g(R)\leq m.
\]
If $c_1,c_2,c_3$ are distinct, consider the block $B(c_1,c_2,c_3)$. An
item of this block assigned to buyer $i\leq 3$ lies outside $A_i(c_i)$
and contributes zero to the sum. If none of the block's items is
assigned, the whole block remains and $g(R)=|R|-1$. In either case,
conditional expected welfare is at most $m-1$.

Buyer 2's row avoids $c_1$ with probability $1-1/a$. Given the history
through buyer 2 on that event, buyer 3's independent column avoids both
cells with probability at least $1-2/b$. Thus all three cells are
distinct with probability at least $p_{a,b}=(1-1/a)(1-2/b)$, giving
$\E[\ALG]\leq m-p_{a,b}$.

\paragraph{A matching online rule.}
Buyer 1 uses $(1,1)$. Buyer 2 uses $(1,1)$ if $I=1$ and $(I,2)$ otherwise.
Buyer 3 reuses an earlier cell in her column if possible, and chooses
any cell in her column otherwise. Buyer $i$ receives every remaining
item in $A_i(c_i)$, earning one per item. The final bundle $R$ consists of
the items with $x(c_i)\ne i$ for all three buyers.

If the cells are distinct, $R$ is a block. If buyers $i,j$ share a cell
$u$, then every block $B(e)$ contains an item outside $R$. To construct
one, choose a label in $\{i,j\}$ at $u$ allowed by $B(e)$, which forbids
at most one label there. Put $2$ at $e_1$ if $e_1\ne u$, and put $1$ at
all remaining cells. Thus $d(R)=1$ exactly when the cells are distinct.

Giving $R$ to buyer 4 yields conditional expected welfare $m-d(R)$.
The cells are distinct exactly when $I\ne 1$ and $J\notin\{1,2\}$,
which has probability $p_{a,b}$. The rule therefore attains
$m-p_{a,b}$, and \eqref{eq:prophet} gives the optimal online ratio
$r_{a,b}(q)=(m-p_{a,b})/(2m-1-qm)$.

\paragraph{Choosing the grid dimensions.}
For a fixed grid, $r_{a,b}(q)$ increases with $q$, so its infimum is
\[
  \lim_{q\downarrow 0}r_{a,b}(q)
  =\frac12-\frac{p_{a,b}-1/2}{1+4(ab)^2}.
\]
Maximizing the subtracted term over integers $a,b\geq 2$ gives the
largest limiting gap. The choice $a=4$, $b=9$ yields $m=2{,}593$ and
$p_{4,9}=7/12$, for a limiting ratio of $1/2-1/62{,}220$. Taking
$q=1/(54m)$ gives the exact ratio
\[
  r_{4,9}\left(\frac{1}{54m}\right)
  =\frac{279{,}981}{559{,}978}
  =\frac12-\frac{4}{279{,}989}.
\]
This proves \cref{thm:main} with $\varepsilon=4/279{,}989$.

\section{At most three buyers}
\label{sec:3-buyers}

\begin{theorem}
With at most three buyers whose valuations are independent, monotone,
and subadditive, there is an online allocation rule satisfying
$\E[\ALG]\geq \tfrac12\E[\OPT]$.
\end{theorem}
\begin{proof}
With one buyer, allocating $M$ is optimal. With two buyers,
$\OPT\leq v_1(M)+v_2(M)$ by monotonicity. Allocating $M$ to the buyer with
larger $\E[v_i(M)]$ therefore gives a $1/2$ guarantee.

For three buyers, let $f(R)=\E[v_3(R)]$. Before any buyer arrives,
select one of buyers 1 and 2 uniformly at random and skip the other.
On the selected buyer $i$'s arrival, give her a bundle $B\subseteq M$
maximizing $v_i(B)+f(M\setminus B)$. Give the remaining items to buyer 3.

Fix $v_1,v_2$ and let $(S,T)$ maximize $v_1(S)+v_2(T)$ over partitions of
$M$. By independence, conditional on $v_1,v_2$ and the selection, buyer
3 has expected value $f(R)$ for the remaining bundle $R$. Each of the
first two buyers is selected with probability $1/2$. Comparing their
chosen bundles with $S$ and $T$, respectively, gives
\[
\begin{aligned}
  2\E[\ALG\mid v_1,v_2]
  &\geq v_1(S)+f(T)+v_2(T)+f(S)\\
  &\geq v_1(S)+v_2(T)+f(M),
\end{aligned}
\]
where the second inequality uses subadditivity of $f$. The prophet's bundles
for the first two buyers can be extended to a partition of $M$ without
decreasing their value. By the choice of $(S,T)$, those buyers therefore
contribute at most $v_1(S)+v_2(T)$. The third contributes at most
$v_3(M)$, whose conditional expectation is $f(M)$ by independence.
Thus $2\E[\ALG\mid v_1,v_2]\geq\E[\OPT\mid v_1,v_2]$.
Averaging over $v_1,v_2$ yields $\E[\ALG]\geq\tfrac12\E[\OPT]$.
\end{proof}

\paragraph{Use of AI.}
While developing the results in this note, we worked interactively with
OpenAI's GPT-5.6 Sol, GPT-6 Ultra, and GPT-6.2 Ultra. Over several rounds
of discussion, we gave the models our own ideas and candidate
constructions to explore. We manually verified and simplified the
four-buyer construction to make its main idea easier to follow. We take
full responsibility for the contents of this paper.

\bibliographystyle{plain}
\bibliography{ref}
\end{document}